# Multi-watt long-wavelength infrared femtosecond lasers and resonant enamel ablation

*Xuemei Yang[1,7], Dunxiang Zhang[1,7], Weizhe Wang[1], Kan Tian[2], Linzhen He[1], Jinmiao Guo[1], Bo Hu[1], Tao Pu[1], Wenlong Li[3], Shiran Sun[4], Chunmei Ding[4], Han Wu[1], Kenkai Li[5], Yujie Peng[5], Jianshu Li[4], Yuxin Leng[5,6], and Houkun Liang[1,\*]*

[1] *School of Electronics and Information Engineering, Sichuan University, Chengdu, Sichuan 610064, China*

[2] *Hymson Laser Intelligent Equipment (Chengdu) Co., Ltd., Chengdu 641419, China*

[3] *Chengdu Dien PHOTOELECTRIC Technology Co., Ltd. Chengdu, Sichuan 610100, China*

[4] *College of Polymer Science and Engineering, Sichuan University, Chengdu, Sichuan 610065, China*

[5] *State Key Laboratory of High Field Laser Physics and CAS Center for Excellence in Ultra-intense Laser Science, Shanghai Institute of Optics and Fine Mechanics (SIOM), Chinese Academy of Sciences (CAS), Shanghai 201800, China*

[6] *lengyuxin@siom.ac.cn*

[7] *These authors contributed equally to this article*

\* *hkliang@scu.edu.cn*

High-power broadband tunable long-wavelength infrared (LWIR) femtosecond lasers operating at fingerprint wavelengths of 7-14 μm hold significant promise across a range of applications, including molecular hyperspectral imaging, strong-field light-matter interaction, and resonant tissue ablation. Here we present 6-12 μm broadband tunable parametric amplifier based on $LiGaS_2$ or $BaGa_4S_7$, generating new record output power of 2.4 W at 7.5 μm, and 1.5 W at 9.5 μm, pumped by a simple and effective thin-square-rod Yb:YAG amplifier producing 110 W 274 fs output pulses. As a proof of concept, we showcase efficient resonant ablation and microstructure fabrication on enamel at the hydroxyapatite resonant wavelength of 9.5 μm, with a laser intensity two orders-of-magnitude lower than that required by non-resonant femtosecond lasers, which could foster more precision surgical applications with superior biosafety.

## 1. Introduction

The development of broadband tunable, femtosecond long-wavelength infrared (LWIR) lasers, covering the fingerprint region from 7 to 14 μm, containing complex bending and stretching molecular resonant absorption lines, has spurred extensive research into spectroscopic applications, such as molecular sensing[1] and field-resolved spectroscopy of biological tissues[2]. With the power scaling of LWIR femtosecond lasers, new frontiers in two-dimensional infrared spectroscopy and hyperspectral imaging,[3] efficient high-harmonics generation in solids,[4] as well as minimally invasive tissue ablation[5] have been unlocked. Over the past decade, high-power LWIR femtosecond lasers have undergone rapid development, achieving power scaling from a few to hundreds of milliwatts.[6-19] The technique of intra-pulse difference-frequency generation (IPDFG) could produce multi-octave spanning LWIR spectrum, covering the entire fingerprints regime at a wavelength window of 6-20 μm.[6-11] Notably, a 500 mW LWIR generation through IPDFG has been demonstrated, employing a thulium-doped fiber chirped-pulse amplifier (CPA) operating at 2 μm with 30 W average power and 32 fs pulse width[11]. Optical parametric chirped-pulse amplification (OPCPA) represents a superior pathway to accessing high-energy LWIR femtosecond pulses.[12-14] 0.75 mJ, 7 μm pulses with an average power of 75 mW have been generated through OPCPA, pumped by a 2 μm high-energy Ho:YLF amplifier.[13] Leveraging high-power Yb:YAG CPA, 9 μm OPCPA with 140 mW average power has been achieved based on large-bandgap $LiGaS_2$ (LGS) crystals.[14] Additionally, a number of high-power LWIR femtosecond lasers extending wavelengths beyond 6 μm have been successfully realized through optical parametric amplifier (OPA), utilizing various non-oxide nonlinear crystals, such as $CdSiP_2$,[15] LGS,[16,17] $BaGa_4Se_7$ (BGSe),[18] and $BaGa_4S_7$ (BGS).[19] In a notable demonstration in 2018, an 8.5 μm OPA with 1.3 W output power, the highest reported power of LWIR femtosecond lasers within the fingerprint regime to the best of our knowledge, is demonstrated based on a LGS-OPA pumped by a 50-W Yb:YAG thin-disk oscillator.[17] However, further power scaling towards multi-watt output, especially for wavelengths coinciding with Amide I, Amide II and phosphate resonances at 6.1, 6.4 and 9.5 μm respectively, has yet to be realized, which is under a great demand by more advanced applications, such as femtosecond laser assisted minimally-invasive therapy and surgery.

With the progression of wide-bandgap nonlinear crystals, such as LGS, $LiGaSe_2$, BGS and BGSe, high-power 1 μm lasers emitting pulses shorter than 300 fs have emerged as primary pump sources for high-power LWIR femtosecond OPA systems. Fiber amplifiers have come forth as one of the best choices for delivering high average power and short pulse widths typically between 200 and 400 fs. Pedersen et al. demonstrated a photonic crystal rod fiber-based CPA system delivering 357 fs pulses at 1030 nm with an average power of 175 W at a repetition rate of 750 kHz.[20] To further boost the output power which is constrained by the small core size of the photonic crystal rod fiber, techniques such as thin-rod,[21-23] Innoslab[24] and thin-disk amplifiers[25] have been explored for their remarkable thermal management and beam quality control. Wang et al. investigated a high-power and high-efficiency thin-rod Yb:YAG amplifier comprising three stages of thin-rod amplifications, achieving 200 W output pulses with a duration of 774 fs.[22] Schmidt et al. reported an Innoslab amplifier producing an average power of 540 W and a pulse width of 1.5 ps.[24] Remarkable output power exceeding 1 kW with a pulse duration of 1.1 ps has been demonstrated in thin-disk amplifiers.[25] However, implementing amplification modules in thin-rod, Innoslab, and thin-disk amplifiers requires intricate heat sinks and sophisticated laser head engineering. Furthermore, achieving pulse widths shorter than 300 fs is typically necessary for efficient femtosecond parametric amplification. Therefore, a high-power amplifier capable of generating short pulses with simple gain modules is highly sought after to drive the power scaling of LWIR OPA systems and enable the pursuit of corresponding applications.

In this work, we propose and demonstrate a thin-square-rod Yb:YAG amplifier within a simple laser modular configuration designed for efficacious thermal management and suppression of gain narrowing. Direct output of 110 W, 274 fs 1030 nm pump pulses at a repetition rate of 500 kHz is achieved. Leveraging the high-power pump source, 6-12 μm broadband tunable femtosecond laser emission is generated through 2-stage OPAs utilizing LGS or BGS crystals, achieving record-high output powers of 2.4 W at 7.5 μm and 1.5 W at 9.5 μm. As a proof-of-concept, efficient enamel ablation by the LWIR femtosecond laser tuned to the phosphate resonance wavelength of 9.5 μm is demonstrated, reducing the required laser intensity by two-orders of magnitude compared to off-resonant femtosecond lasers. To the best of our knowledge, this represents the first report of enamel ablation utilizing a LWIR femtosecond laser at the hydroxyapatite vibration resonating wavelength, which is promising for applications such as caries removal and tooth sculpting for crown or filing preparation, and enamel treatment for enhancing the acid and corrosion resistance. Additionally, enamel microstructures are fabricated to enhance the tooth surface roughness and augment the bond strength of orthodontic brackets to enamel.

## 2. Experimental Setup

The high-power femtosecond Yb-doped CPA laser system is composed of a high-power seed source, two-stage water-cooled thin-square-rod Yb:YAG power amplifiers and a pulse compressor, as illustrated in **Figure 1**(a). The seed source, a customized fiber laser amplifier (YACTO-FL50), provides a train of stretched pulses with a maximum average power of 50 W and a pulse width of 400 ps at a repetition rate of 500 kHz. The seed spectrum is centered at 1032 nm with a full width at half-maximum of 9 nm. A 1-at% doped thin-square-rod Yb:YAG crystal with 20 and 15 mm in length and additional 2-mm-long undoped endcaps on both sides, as depicted inset of **Figure 1**(a), is mounted in a water-cooled heat sink with four surrounding cooling surfaces in the first and second-stage amplifiers, respectively. The beam diameters of seed lasers on Yb:YAG crystals in two-stage amplifiers are designed as 550 and 600 μm, respectively. Both stages are end-pumped by a high brightness fiber-coupled laser diode delivered through a multimode fiber with a diameter of 200 μm, emitting up to 300 W output power at a locked wavelength of 940 nm. The pump beams are coupled into thin-square-rod Yb:YAG crystals by lens pairs, with corresponding pump beam waists of 600 and 770 μm in the first and second-stage amplifiers, respectively. Subsequently, the amplified pulse is injected into a compressor comprising a transmission grating pair.

Driven by the high-power femtosecond CPA utilizing thin-square-rod Yb:YAG crystals, a multi-watt broadband tunable LWIR OPA is experimentally demonstrated, illustrated by the schematics in **Figure 1**(b). The compressed pulse from the Yb:YAG CPA exhibits an average power of 110 W, and a 3-W pump power is split and focused into a 10-mm-thick YAG crystal to generate a stable supercontinuum (SC) via single filamentation. Subsequently, the near-infrared band of SC is selected after passing through a 1100-nm long-pass filter, yielding broadband signal pulses of LWIR OPA, with an average power of 33 mW and extending up to a wavelength of 1800-nm. In the first OPA stage, an uncoated 8-mm-thick LGS crystal, cut for type-I phase matching is employed, positioned within a diverging pump beam with a diameter of ~ 800 μm. The amplified signal, driven at a 10-W pump power, is tunable from 1.13 to 1.28 μm, with a power varying in the range of 0.3 to 1 W.

The amplified signal pulse proceeds into the second-stage OPA, driven by the remaining 87-W pump power. Here, an 8-mm-thick LGS crystal, cut for type-II phase matching, which provides a 20%-greater nonlinearity compared to its type-I counterpart is employed. The pump and signal beams are directed onto the LGS crystal with diameters of ~1.8 mm and ~1.6 mm, respectively, with the maximum pump intensity reaching 22 $GW/cm^2$ on the front surface of LGS crystal. To scale power towards longer wavelengths, the LGS crystal is replaced by a 10-mm-thick uncoated BGS crystal (DIEN TECH), cut for type-I phase matching (theta=11.7°). The generated LWIR emission is collected by a hollow core mid-infrared fiber (OptpKnowledge HF500MW) and characterized by a scanning-grating monochromator (Zolix Omini-λ500i) equipped with a lock-in amplifier (SR830) and a liquid-nitrogen-cooled mercury cadmium telluride detector (Judson, DMCT16-De01).

As a proof of concept, enamel femtosecond resonant ablation is performed at a wavelength of 9.5 μm, coinciding with the phosphate absorption peak. Extracted caries-free cow teeth (all work conducted under ethical approval) are sectioned with a diamond saw into 1-2 cm slices and polished with a SiC #600 and #1200 grinding papers, providing well defined flat enamel surfaces and mirror finishing. These tooth slices are affixed on an electrically-driven 2-dimensional translation platform, as depicted in **Figure 1**(c). The LWIR laser beam is subsequently directed towards the tooth slices and focused through a ZnSe lens with a focal length of 50 mm, resulting in a focal spot diameter of 90 μm. A scanning ablation technique is employed, with a scanning speed set at 0.5 mm/s.

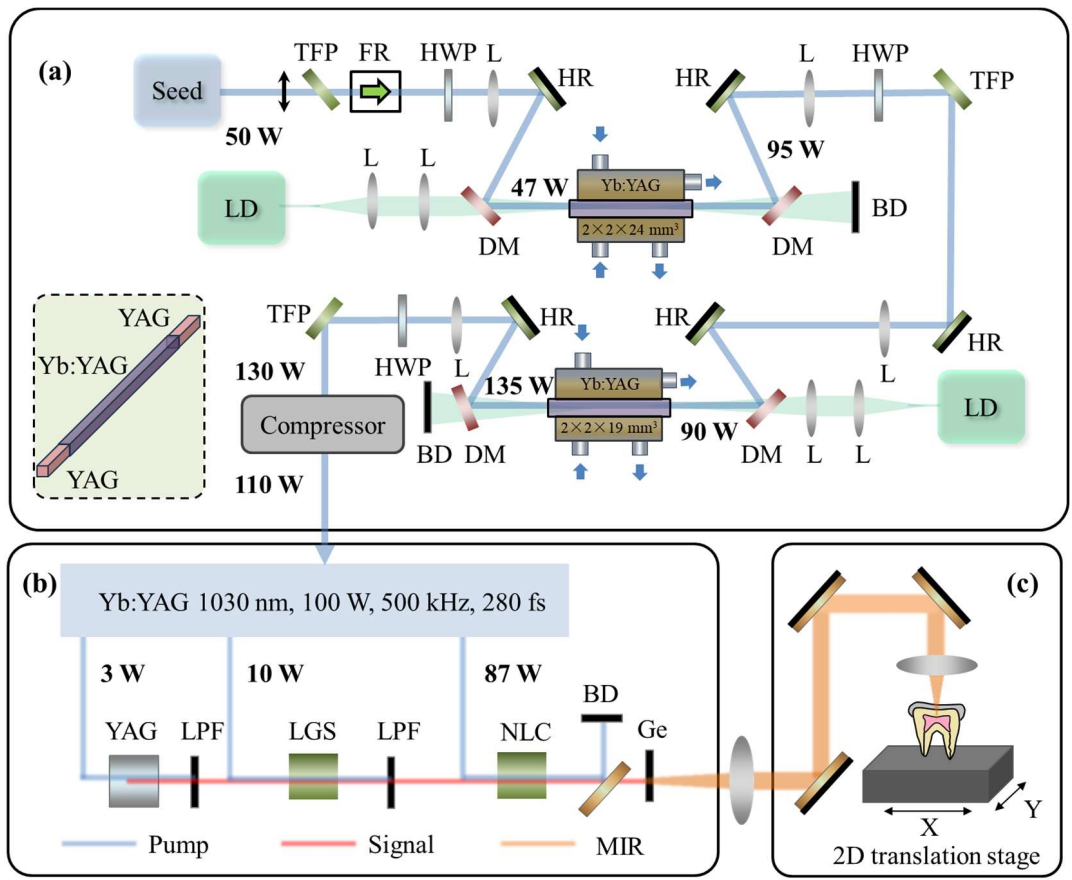

**Figure 1. Schematics of overall experimental setup.** (a) The schematic of thin-square-rod Yb:YAG chirped-pulse amplifier (CPA) system. The CPA consists of a 50-W, 1030 nm fiber laser amplifier as a high-power seed source, two-stage CPA driven by two 300-W, 940 nm laser diodes, and a customized pulse compressor. Output powers at different stages are marked. The inset depicts the thin-square-rod Yb:YAG crystal with 2-mm-long undoped endcaps on both sides. Yb:YAG crystals are mounted in water-cooled heat sinks with four surrounding cooling surfaces. The water flows of the heatsink are also marked by blue arrows. LD, laser diode; HR, high-reflection mirror; TFP, thin film polarizer; HWP, half wave plate; L, lens; FR, Faraday rotator; DM, dichroic mirror; BD, beam dump. (b) The schematic of the high-power long-wavelength infrared (LWIR) broadband tunable femtosecond optical parametric amplifiers; LPF, long-pass filter; NLC, nonlinear crystal, LGS or BGS; Ge, germanium long-pass filter. (c) The schematic of enamel resonant ablation by a LWIR femtosecond laser platform.

## 3. Results
### 3.1. 110-W, sub-300 fs, thin-square-rod Yb:YAG CPA

An efficacious thermal management of laser amplifier is crucial for power scaling and beam quality control. To investigate the thermal distribution of the thin-square-rod Yb:YAG crystal mounted in a customized water-cooled heat sink, the thermal simulation is performed using a finite-element tool as illustrated in **Figure 2**(a). A thin-square-rod Yb:YAG crystal with a cross-section of 2 mm×2 mm, and a length of 20 mm is firstly investigated. It is assumed that the absorbed pump energy in the crystal has a uniform Gaussian distribution over the entire pump volume with a diameter of 1 mm at the entrance of Yb:YAG crystal, ensuring the simulation results close to the experimental conditions. With 300 W pump operating at 940 nm, 30 W (~ 12.5% of total absorbed pump power) heat power is generated in the crystal. The simulation reveals a peak temperature of 48.2 °C at the center of the pump surface, exhibiting a temperature gradient ($\Delta T$) of 32.2 °C between the hottest spot and the cooled face. There is no noticeable distortion in the temperature distribution at corners of the crystal, indicating remarkable heat dissipation capability and thermal management of the thin-square-rod crystal. For comparison, the temperature distribution of a thin-rod Yb:YAG crystal with a 2-mm diameter is also simulated. The results show a maximum temperature of 46.3 °C, with only a slight 1.9 °C reduction compared to the thin-square-rod crystal. The similar temperature distribution of two geometric types suggests that the thin-square-rod crystal has the particular advantage for its simpler heat sink design and fabrication. To further explore this, thermal distribution in a Yb:YAG crystal shaped as a thin-square-rod with a cross-section of 5 mm×5 mm is simulated too. It is observed that the thermal gradient is higher by 17.5 °C compared to the thin-square-rod crystal with a cross-section of 2 mm×2 mm, which highlights the efficacy of good thermal management in thin-square-rod crystals with small cross-sections and large surface-to-volume ratios. Additionally, it is simulated that further reducing the cross-section to 1.5 mm×1.5 mm and a diameter of 1.5 mm decreases the maximum temperature in the Yb:YAG crystals to 41.7 °C and 40.1 °C, respectively. However, this also increases the difficulty in coupling and confining multimode pump beams. Therefore, in this work, a thin-square-rod Yb:YAG crystal with 2 mm×2 mm cross-sectional size is chosen for the optimal balance between effective thermal management and ease of pump beam manipulation.

The balance between a high amplification factor and a short pulse width is another critical criterion of a femtosecond laser amplifier serving as a pump source for an OPA. Prior to experimentation, a simulation is conducted to examine the evolution of the amplified output spectral bandwidth in two-stage Yb:YAG amplifier with varying amplification factors. The

CPA process is modeled by solving the rate equation of the quasi-three level system using the Euler method and the 4th-order Runge-Kutta algorithm. Results indicate that the amplified spectral bandwidth decreases from 9 to 2 nm as the amplification factor increases from 1 to 10, owing to the gain narrowing effect as depicted in **Figure 2**(b). It is noteworthy that selecting an amplification factor of 3 yields an amplified spectrum with a bandwidth of 4.7 nm, supporting a Fourier-transform-limited (TL) pulse width of ~ 210 fs, as presented in Supplementary Figure S1, which is suitable as a pump pulse of an OPA. This suggests that a moderate gain combined with a high-power seed could provide a solution for generating high-power pulses with suppressed gain narrowing for pumping the OPA.

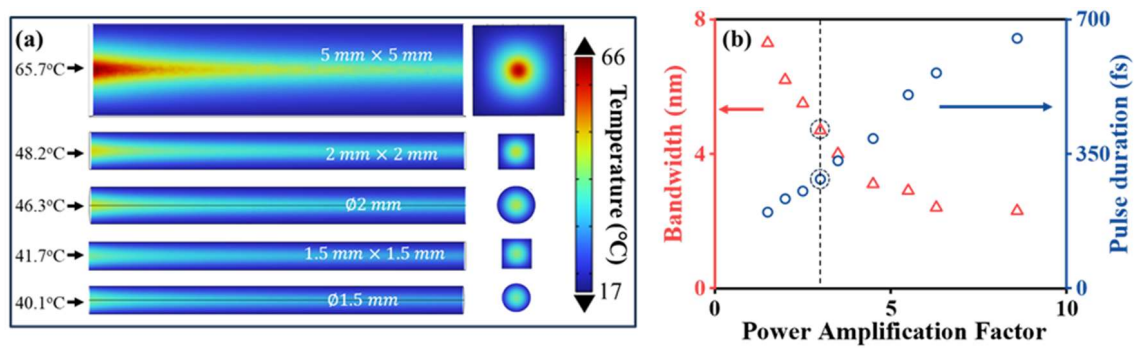

**Figure 2. Modelling of high-power, sub-300 fs, thin-square-rod Yb:YAG chirped-pulse amplifier.** (a) Simulated thermal distribution of Yb:YAG crystals mounted in a customized water-cooled heat sink. Thin-square-rod and thin-rod crystal geometries with varying cross-sectional sizes are compared. Maximum heat power considered in simulation is 30 W (~ 12.5% of total absorbed pump power). The peak temperature occurring at the center of the pump surface in thin-square-rod crystals with a cross-section of 5 mm×5 mm, 2 mm×2 mm and 1.5 mm×1.5 mm, and thin-rod crystals with a diameter of 2 mm and 1.5 mm are simulated as 65.7 °C, 48.2 °C, 41.7 °C, 46.3 °C, and 40.1 °C, respectively. Undoped YAG caps are not displayed, since they do not absorb pump power. (b) Modelling of spectral bandwidth evolution (red triangles) and corresponding Fourier-transform-limited pulse duration (blue circles) as a function of the amplification factor ranging from 1 to 10.

Through the thermal engineering and amplification design, a high-power, sub-300 fs CPA utilizing water-cooled thin-square-rod Yb:YAG crystals are experimentally demonstrated. As characterized in **Figure 3**(a) with blue circles, the output power in the first-stage Yb:YAG amplifier reaches 95 W, corresponding to an extraction efficiency of 20%, as the absorbed pump power is increased to 240 W, with a strong input seed power of 47 W. Subsequently, in the second-stage CPA, the amplified output power increases from 90 to 135 W when a pump power of 225 W is absorbed, as indicated by red circles in **Figure 3**(a). Notably, extraction saturation does not occur in either amplification stage. Therefore, we anticipate further enhancement of output power and extraction efficiency with higher pump power, building upon the current configuration of the amplifier chain.

Subsequently, the evolution of amplified spectra is systematically characterized. **Figure 3**(b) presents the spectra of the seed (grey), the output of the first (blue) and second-stage amplifiers (red). In comparison to the spectrum of the fiber laser seed, the gain spectrum of the first-stage water-cooled Yb:YAG amplifier is centered at 1030.8 nm with an amplified bandwidth of ~ 6.6 nm. The reduction in output bandwidth is attributed to mild gain narrowing and slight misalignment of gain and seed spectral peaks. The amplified spectral bandwidth is further reduced to 4.5 nm in the second-stage Yb:YAG amplifier, supporting ~ 230 fs TL pulses, as depicted in Supplementary Figure S1. The amplified pulse from the CPA system is directed into a grating compressor, and the temporal profile is charactered when the CPA is operated at maximum output power, using a second-harmonic generation Frequency-Resolved Optical Gating (FROG) technique. The measured and retrieved spectra have a relatively good agreement as presented in **Figure 3**(c), with a bandwidth of 4.5 nm supporting a TL pulse of 230 fs, assuming a Gaussian profile. **Figure 3**(d) showcases the retrieved temporal profile, revealing a pulse width of 274 fs. The unsuppressed pedestals, holding ~ 7% of the total energy from the measured pulse, are associated with the uncompensated third-order dispersion. The measured and retrieved FROG traces of the 274-fs pulse are presented in Supplementary Figure S2.

With a compression efficiency of 84.6%, a compressed power of 110 W is obtained. In a long-term stability measurement as presented in **Figure 3**(e), a standard deviation of 0.3% is measured over a 4-hour period, which fulfills the requirement as a pump source of LWIR OPA with subsequent medical applications. To assess the thermal-optic effects of the high-power water-cooled CPA utilizing thin-square-rod Yb:YAG crystals, the beam quality factor, $M^2$ is characterized. As shown in **Figure 3**(f), the measured $M^2$ values are 1.10 and 1.09 for the horizontal and vertical directions, respectively. In the inset of **Figure 3**(f), a round Gaussian beam is manifested, indicating effective thermal management of the CPA system with water-cooled thin-square-rod Yb:YAG crystals.

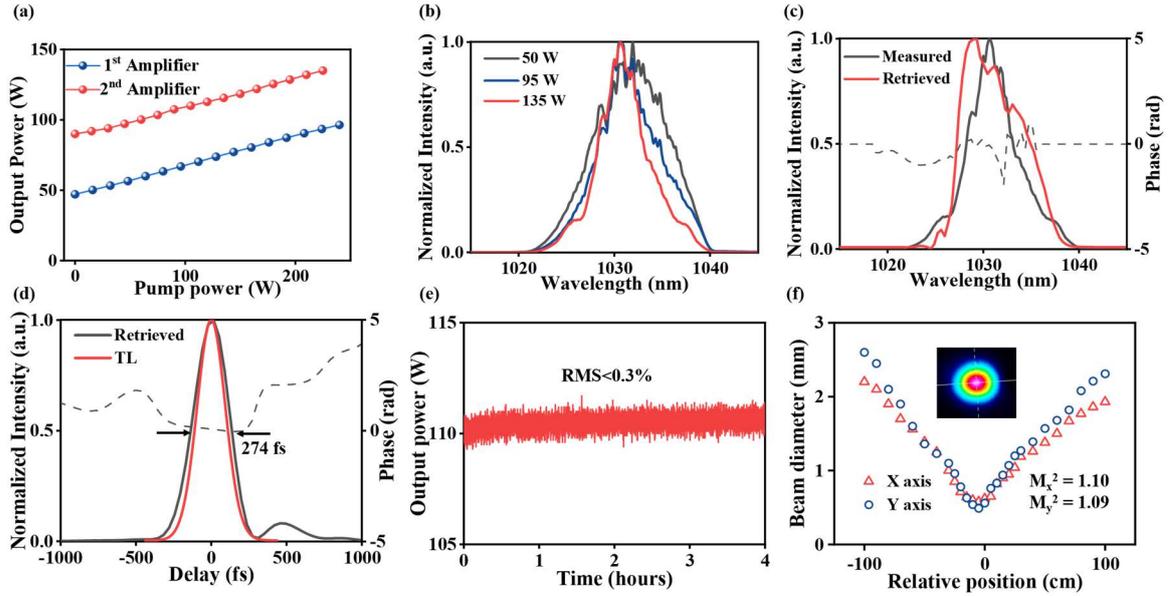

**Figure 3. Characterization of high-power, sub-300 fs, thin-square-rod Yb:YAG chirped-pulse amplifier.** (a) The amplified output power in the first (blue circles) and second-stage (red circles) Yb:YAG amplifiers as a function of the absorbed pump power. (b) Comparison of the spectra from seed source (grey), the first (blue) and second-stage Yb:YAG amplifiers (red). (c) The measured and retrieved Frequency-Resolved Optical Gating (FROG) spectral intensity and phase of the output pulse. (d) The retrieved FROG temporal profile and the transform-limited temporal profile, indicating a 274-fs pulse width. The FROG error is 0.8%. (e) The long-term stability measurement of the compressed output power. A standard deviation of 0.3% is recorded over a 4-hour period. (f) The measured M2 factor of the Yb:YAG laser system operating at full power. The inset shows the near-field image of the output beam profile.

## 3.2. Mutil-watt, 6-12 μm tunable femtosecond LWIR OPA

The demonstrated high-power sub-300-fs Yb:YAG CPA is employed as the pump source for the LWIR OPA. In **Figure 4**(a), experimentally measured idler spectra from the LWIR OPA based on LGS crystals are presented, showcasing a broad tunable wavelength range spanning from 5.5 to 12 μm. The spectral width remains relatively narrow over two spectral ends, but broadens around wavelengths of 7-10 μm, where the phase-matching bandwidth is larger. Further extension of the idler wave to longer wavelengths is hindered by the absorption edge of the LGS crystal at 12 μm. **Figure 4**(b) plots the average power of the generated LWIR femtosecond pulses centered at 6.4 μm (yellow), 7.5 μm (red), 8.5 μm (blue) and 9.5 μm (grey) as a function of the pump power. The highest LWIR output power of 2.4 W at 7.5 μm is achieved under a pump power of 87 W, with a corresponding power conversion efficiency of ~ 2.8% and a quantum efficiency of ~20%. To the best of our knowledge, this is the highest power of LWIR femtosecond lasers in the fingerprint region. Notably, the output power of LWIR emission at 9.5 μm, resonating with the hydroxyapatite vibration, drops abruptly to 0.94 W. Subsequently, to elucidate the temporal characterization of the LWIR output from the femtosecond OPA based on LGS crystals, the typical temporal profile of idler pulses at a central wavelength of 7.5 μm is characterized by using a home-built interferometric autocorrelator (IAC) setup. As presented in **Figure 4**(c), the measured IAC trace with a ratio of 1:8 between the background and the peak of the IAC signal exhibits 12 optical cycles corresponding to a 300-fs pulse duration, which is significantly longer than the TL pulse width (~ 100 fs). This is attributed to the large uncompensated dispersion inherited from long LGS crystals. The inset in **Figure 4**(c) depicts a good Gaussian profile of the LWIR output at 7.5 μm, as measured by a pyroelectric camera (WinCam D-IR-BB).

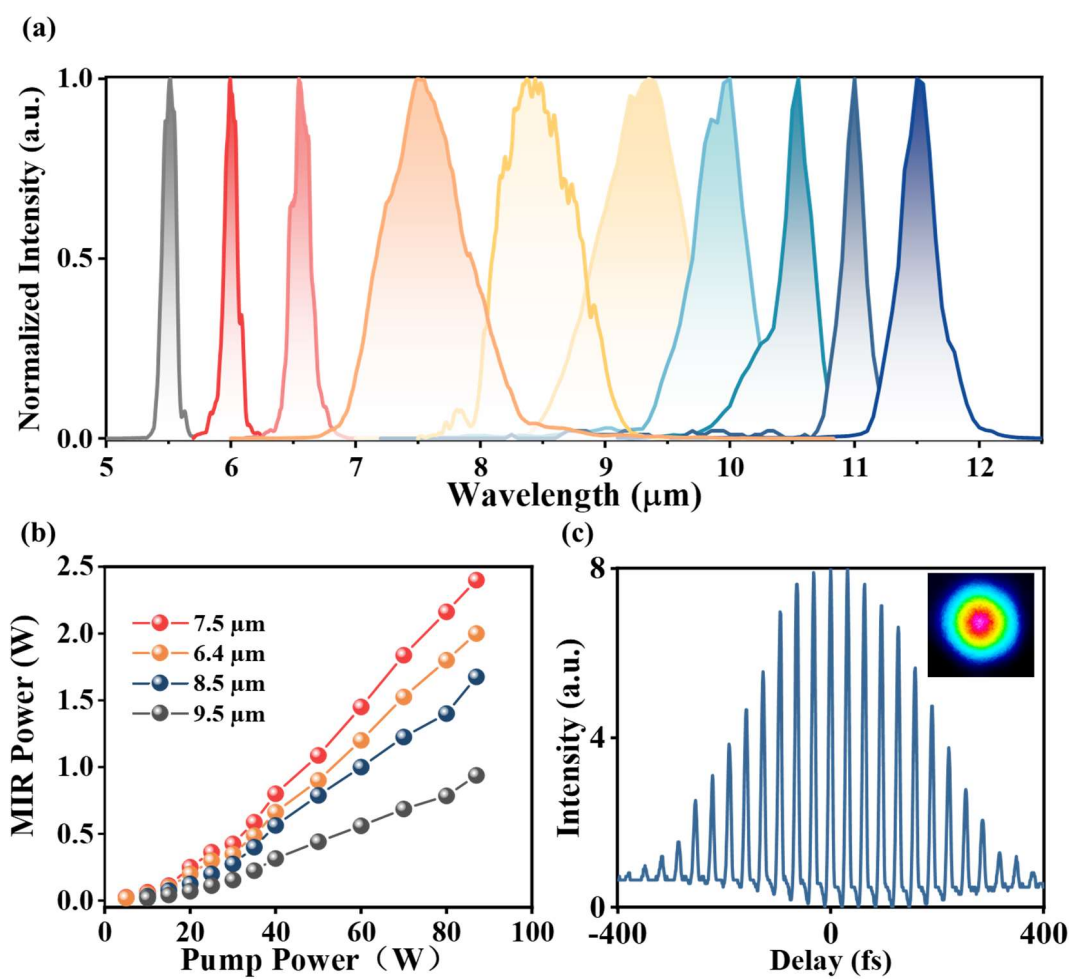

**Figure 4. Multi-watt, broadband tunable, femtosecond, LWIR OPA based on LGS crystals.** (a) Typical spectra of the LGS-OPA with tunable wavelengths spanning from 5.5 to 12 μm. (b) The output power of LWIR pulses with the central wavelength of 6.4 μm, 7.5 μm, 8.5 μm and 9.5 μm as a function of pump power at the second LGS crystal. (c) The measured interferometric autocorrelation trace of the LWIR OPA pulse centered at 7.5 μm, indicating a typical pulse width of ~300 fs. The inset shows the corresponding beam profile of LWIR output revealing a good Gaussian profile.

To investigate power scaling at longer wavelengths resonating with the hydroxyapatite vibration mode, the nonlinear crystal in the second-stage LWIR OPA is replaced with a BGS crystal. The pump intensity is slightly reduced to 20 GW/cm$^2$ to prevent two-photon absorption, as BGS possesses a smaller band-gap energy compared to that of LGS (3.54 eV for BGS vs. 3.87 eV for LGS[26]). LWIR output spectra, presented in **Figure 5**(a), exhibit a broad tuning range from 6 to 13 µm. The spectral width significantly broadens at wavelengths of 9 to 11 µm, benefiting from favorable phase-matching conditions, which supports a TL pulse width of ~ 90 fs. **Figure 5**(b) illustrates the recorded idler wave output power at different central wavelengths, demonstrating 1.5 W power at 9.5 µm, pumped at 80 W. Notably, BGS surpasses LGS in the spectral range above ~ 9 µm, representing a unique advantage of BGS.

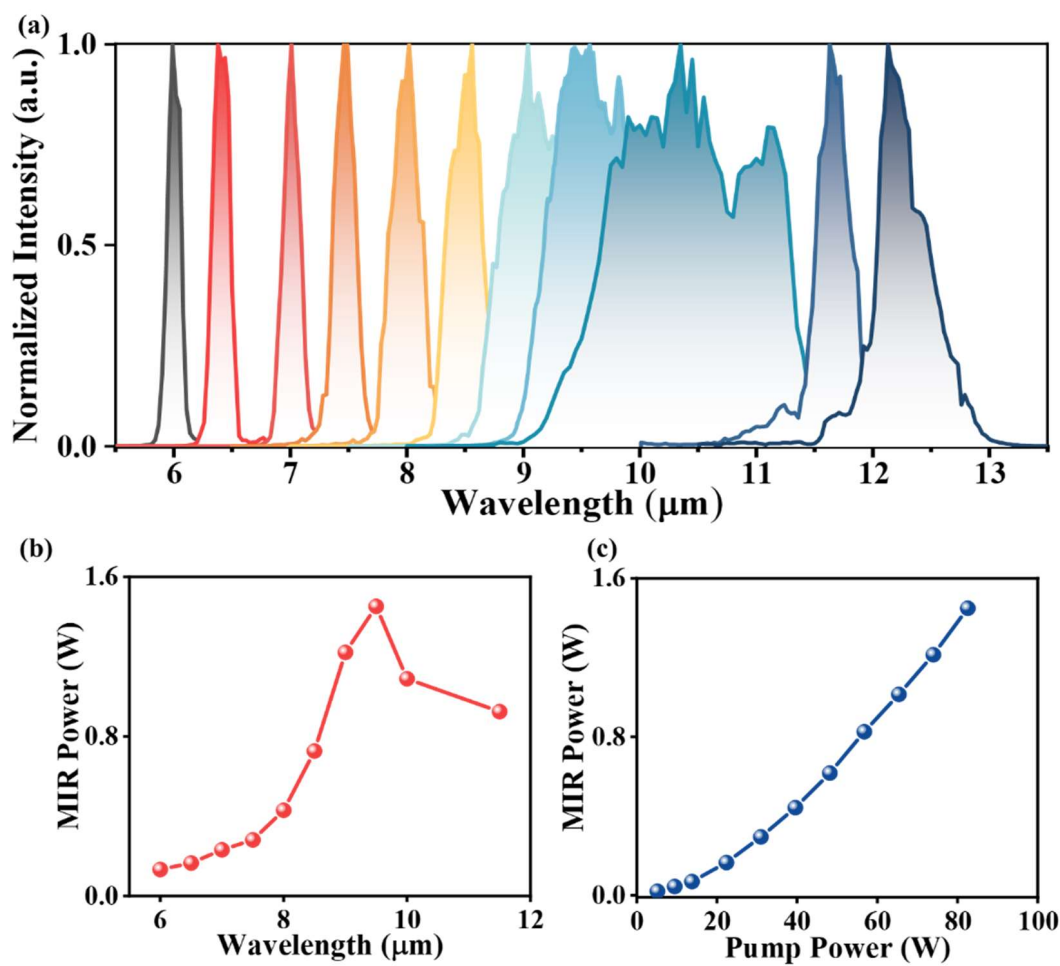

**Figure 5. Multi-watt, broadband tunable, femtosecond, LWIR OPA based on BGS crystal.** (a) Idler spectra measured from the BGS-OPA. (b) Average power as a function of wavelength in BGS-OPA driven by a pump power of 80 W. (c) The output power of the LWIR pulses with the central wavelength of 9.5 µm as a function of pump power at the BGS crystal.

### 3.3. Efficient enamel ablation at phosphate resonance wavelength

It is triggered by the fact that at the LWIR resonant wavelength, laser tissue ablation exhibits superior efficiency and cellular-scale collateral damage.[5] Hence, resonant enamel ablation is performed based on the demonstrated multi-watt LWIR femtosecond OPA. Fourier transform infrared spectroscopy (FTIR) of an enamel sample is firstly conducted to pinpoint the resonant wavelengths within the spectral region of 2-14 μm. As shown in **Figure 6**(a), the most prominent resonant peak at 9.5 μm corresponds to stretching vibrations in phosphate. O-H in water and amide I, II proteins are depicted by FTIR peaks at 3.7 and 6.1-6.4 μm, respectively. Accordingly, the wavelength of LWIR OPA is tuned to 9.5 μm to align with the phosphate resonance in enamel for effective ablation. Pulses of 0.6 W at 9.5 μm are delivered to the enamel surface with some reflection losses. The ablation depths are characterized by a 3-dimensional surface profiler, revealing trenches with depths of 200-240 μm, as presented in **Figure 6**(b). To the best of our knowledge, this marks the first report of enamel ablation through femtosecond LWIR laser at the hydroxyapatite vibration resonating wavelength. It is worth noting that laser intensity at 9.5 μm is ~ 50 GW/cm$^2$, which is two-orders of magnitude smaller than those of other femtosecond lasers at near-infrared off-resonant wavelengths.[27, 28] This guarantees efficient and safe ablation while circumventing tissue ionization, by employing resonant LWIR lasers with small photon energy (0.14 eV). Representative scanning electron microscope images of an ablated enamel trench are depicted in **Figure 6**(c, d), showcasing an adequately rough incision that improves adhesion strength of fillings and enhances the acid and corrosion resistance of the tooth.[29] This opens avenues for potential applications in caries removal and tooth sculpting for crown or filing preparation. Additionally, typical microstructures such as holes, channels, and meshes are ablated utilizing 9.5 μm femtosecond pulses, as shown in **Figures 6**(e-g), respectively, to further enhance the roughness of the tooth surface and augment the bond strength of orthodontic brackets to enamel.[30]

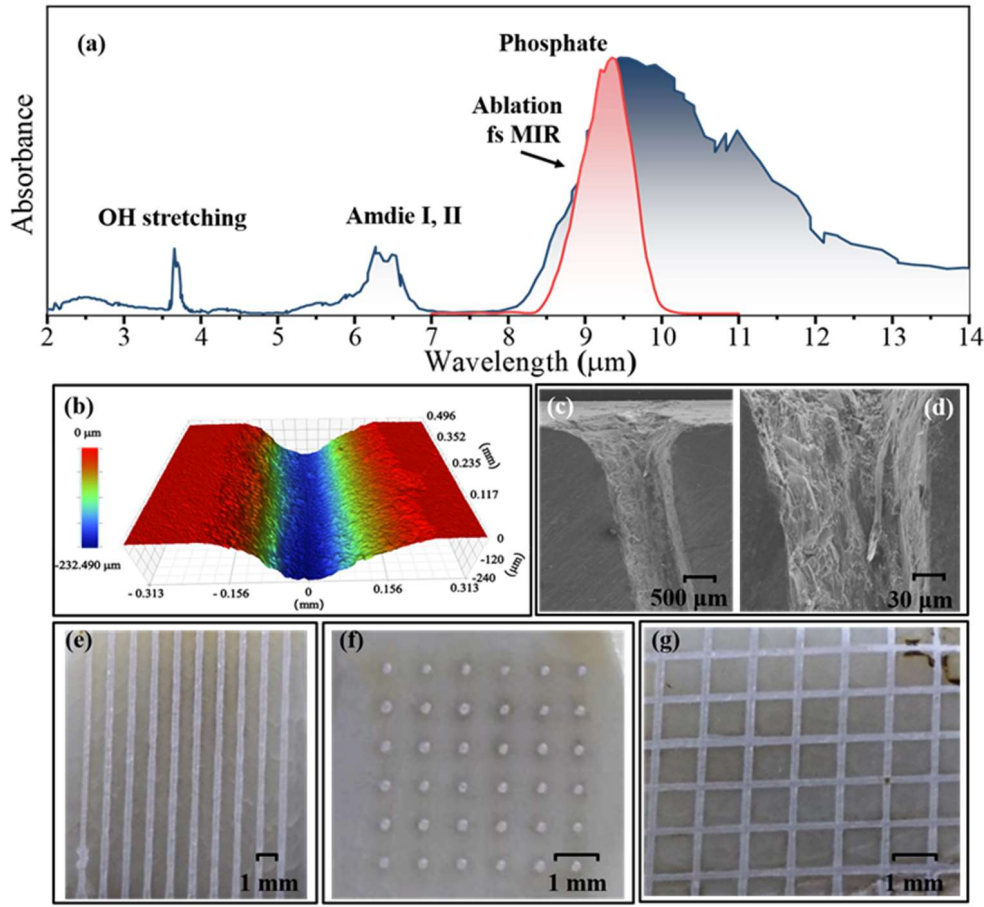

**Figure 6. Enamel resonant ablation at phosphate resonance wavelength of 9.5 μm.** (a) Fourier transform infrared (FTIR) absorption spectrum of enamel in the wavelength range spanning from 2 to 14 μm. (b) shows the 3-dimensional surface profile of ablated enamel tissue with a trench on the tooth slice. A representative scanning electron microscope (SEM) image (c) and the corresponding enlarged image (d) of the enamel incision showing an adequate roughness. (e-g) SEM images of microstructures such as holes, channels, and meshes, respectively, on surfaces of tooth slices, ablated utilizing 9.5 μm femtosecond pulses.

## 4. Discussion and Conclusion

The demonstrated thin-square-rod amplifier provides a simple technique for generating laser power exceeding 100 W with pulse widths below 300 fs. Further scaling up to 200 W is viable by reducing the rod cross-section to 1 mm×1 mm, which could additionally decrease the temperature gradient by ~ 15 °C, as illustrated in the thermal simulation in Supplementary **Figure S3**. This suggests that the thin-square-rod amplifier can achieve comparable thermal management and power handling capabilities, albeit with a simpler design of gain module and heat sink, in comparison to the thin-rod amplifier. Therefore, linear scaling of LWIR power at fingerprints wavelengths over 5 W is foreseeable. In the efficient enamel ablation demonstrated at the phosphate resonant wavelength of 9.5 μm, significantly less laser intensity is required compared to non-resonant wavelengths, ensuring superior biosafety, aided by femtosecond cold processing. Further power scaling to the 5-W level could meet the demand for routine teeth drilling and reduce patient discomfort resulting from turbodrills. Moreover, multi-watt broadband tunable fingerprint lasers, aligned with individual main resonant peaks in bio-tissues could unlock other clinical research avenues such as selective pancreas tumor ablation, atheromatous plaque ablation, and multi-micro-channel formation in cartilage. Furthermore, with the advancement of LWIR wide-bandgap nonlinear crystals, such as LGS, BGS and BGSe, accessing high-power LWIR pulses spanning from 6 to 16 μm could become easier, which could open up more possibilities in applications such as long-range molecular sensing in the atmospheric transmission window of 8-14 μm, efficient terahertz generation and high-flux high-harmonic generation in solids.

In summary, we have developed a high-power Yb:YAG amplifier utilizing thin-square-rod crystals to ensure effective thermal management. Combined with the optimized gain design, 110 W pulses at a 500 kHz repetition rate with a pulse duration of 274 fs are realized. Pumped by the high-power femtosecond laser, LWIR OPA with broadband tunable emission in a spectral range spanning from 6-12 μm is generated, achieving output of 2.4 W at 7.5 μm and 1.5 W at 9.5 μm which are new records in the fingerprint region, to the best of our knowledge. Additionally, we have conducted a proof-of-concept demonstration showcasing efficient enamel ablation by tuning the high-power LWIR femtosecond OPA to a wavelength resonating with the hydroxyapatite vibration mode at 9.5 μm. Notably, femtosecond resonant ablation significantly reduces the required intensity by two-orders of magnitude compared to the near-infrared non-resonant femtosecond lasers. This breakthrough paves the way for laser-assisted dental caries treatment and root canal therapy, offering a promising avenue for improving oral healthcare practices.

**Supporting Information.** The data that support the findings of this study are available from the corresponding author upon request.

**Acknowledgements.** The authors gratefully acknowledge Analytical & Testing Center, Sichuan University for their help in FTIR measurement.

**Funding.** National Natural Science Foundation of China (62075144, U22A2090, U22A20158), Sichuan Outstanding Youth Science and Technology Talents (2022JDJQ0031), Engineering Featured Team Fund of Sichuan University (2020SCUNG105).

**Supplemental document.** See Supplementary Information for supporting content.

**Conflict of Interest** The authors declare no conflicts of interest.